\documentclass[fleqn]{2017SCGE}
\begin{document}
\ensubject{subject}
\ArticleType{Article}
\SpecialTopic{SPECIAL TOPIC: }
\Year{2018}
\Month{***}
\Vol{***}
\No{***}
\DOI{*******}
\ArtNo{000000}
\ReceiveDate{January 11, 2016}
\AcceptDate{April 6, 2016}

\title{The effect of Lorentz-like force on collective flows of K$^{+}$ in Au+Au collisions at 1.5 GeV/nucleon}{The effect of Lorentz-like force on collective flows of $K^{+}$ in Au+Au collisions at 1.5 GeV/nucleon}
\author[1,2]{Yushan Du}{}
\author[1]{Yongjia Wang}{wangyongjia@zjhu.edu.cn}
\author[1,3]{Qingfeng Li}{liqf@zjhu.edu.cn}
\author[2]{Ling Liu}{}

\AuthorMark{Du Y S}

\AuthorCitation{Du Y S, Wang Y J, Li Q F, et al}

\address[{\rm1}]{School of Science, Huzhou University, Huzhou 313000, China}
\address[{\rm2}]{School of Physics science and Technology, Shenyang Normal University, ShenYang 110034, China}
\address[{\rm3}]{Institute of Modern Physics, Chinese Academy of Sciences, Lanzhou 730000, China}


\abstract{Producing kaon mesons in heavy-ion collisions at beam energies below their threshold energy is an important way to investigate the properties of dense nuclear matter.
In this study, based on the newly updated version of the ultrarelativistic quantum molecular dynamics model, we introduce the kaon-nucleon (KN) potential, including both the scalar and vector
(also dubbed Lorentz-like) aspects. We revisit the influence of the KN potential on the collective flow of K$^{+}$ mesons produced in Au+Au collisions at $E_{lab}$ = 1.5 GeV/nucleon and find that the contribution of the newly included Lorentz-like force is very important, particulary for describing the directed flow of K$^{+}$.
Finally, the corresponding KaoS data of both directed and elliptic flows can be simultaneously reproduced well.}


\maketitle


\begin{multicols}{2}
\section{Introduction}\label{section1}
The study of the thermodynamic relationship between the energy (or the pressure) and density (or chemical potential) in nuclear matter, i.e., the nuclear equation of state (EoS), has attracted considerable research interest in nuclear physics and astrophysics since it is essential to understand the various phenomena occurring in nuclear structures and reactions, as well as in, for example, neutron stars. Thus far, the EoS of isospin symmetric (having the same number of neutrons and protons) nuclear matter is relatively well understood \cite{Hartnack:2005tr}; however, information regarding isospin asymmetric nuclear matter is lacking, for which a large uncertain source is density-
\Authorfootnote
\noindent dependent nuclear symmetry energy $E_{sym}(\rho)$, especially at supranormal densities (see, e.g., Refs.~\cite{BALi08,Tsang:2012se,Lattimer:2012xj,Xiao:2009zza,Feng:2009am,Russotto:2016ucm,Guo:2012aa,xzg,Zou:2016lpk,gb,zouly}).

It is well known that heavy-ion collisions (HICs) compress nuclei such that a piece of nuclear medium with a density higher than the saturation density can be created. Although this medium is extremely short-lived, it provides the only opportunity to study the properties of dense nuclear matter in laboratories on Earth. Due to its short lifetime, its properties cannot be studied directly. Therefore, a transport model that links the nuclear EoS with experimental observables is a prerequisite.

Furthermore, model simulations have proposed that a positively charged kaon meson K$^+$ can be produced in HICs at energies below the threshold energy (1.58 GeV), referred to as subthreshold production, and can be used as a sensitive probe for the stiffness of the EoS \cite{Aichelin:1986ss}. We also know that the $K^{+}$ meson carries a strange antiquark, is predominantly produced in the early collision stage (high-density phase), and cannot be reabsorbed by others due to the conservation of strangeness. Thus, it may carry useful and clean information about the dense nuclear matter in the early stage. The first measurement of subthreshold $K^{+}$ meson production (from Au + Au collision at 1 GeV/nucleon) was performed at GSI-SIS in 1994 \cite{Miskowiec:1994vj}. Since then, extensive theoretical and experimental studies of subthreshold kaon production have been performed. For example, using the $K^{+}$ yield data produced from C + C and Au + Au collisions, calculations based on both isospin quantum molecular dynamics (IQMD) and Lanzhou quantum molecular dynamics (LQMD) models support a soft EoS \cite{Hartnack:2005tr,Feng:2011dp}.

In the recent two decades, the $E_{sym}(\rho)$ attracted considerable research interest and it was proposed that the yield ratio between K$^{0}$ and K$^{+}$ mesons can also be a useful tool for detecting EoS stiffness \cite{Li:2005zza,Ferini:2006je}. Because the isovector aspect of the EoS is weak compared to the isoscalar aspect, we must carefully consider the kaon-nucleon (KN) potential. It has been predicted by different model calculations \cite{KN} that the KN potential is slightly repulsive and that the antikaon-nucleon ($\bar{K}$N) potential is strongly attractive (another motivation for studying the kaon potential is the existence of kaon condensation in neutron stars \cite{Lattimer:2006xb} and $\bar{K}$ bound nuclei \cite{Akaishi:2002bg}). Experimental comparisons have revealed that the KN potential at normal nuclear density is of the order of 20-30 MeV. However, a large uncertainty regarding the depth of the $\bar{K}$N potential has also been found, i.e., the $\bar{K}$N potential at normal nuclear density varies from -40 MeV to -200 MeV \cite{Schaffner:1996kv,Li:1997zb,Ha06,Fuchs:2005zg}.

With respect to the KN potential, researchers have pointed out that it has both scalar and vector aspects due to its relativistic origin, and a Lorentz-like force (LF) from the spatial component of the vector field must be taken into account in transport model simulations to better describe experimental kaon flow data \cite{Fuchs:1998yy,Wang:1998fpa,Zheng:2002mj,Fuchs:2005zg}. In our previous work,
we also found that the calculated directed flow slope of $K^{+}$ at mid-rapidity becomes too negative when only the scalar aspect of the KN potential is taken into account \cite{LiMiao:2017}. In our current work, we introduce the LF force into the ultrarelativistic quantum molecular dynamics (UrQMD) model, and we revisit its influence on both the directed and elliptic flows of $K^{+}$. In the next section, we show the corresponding model updates. In section\ref{sec:3}, we describe and discuss the contribution of the KN potential, particularly the LF force, to $K^{+}$ flows. Finally, we provide a summary and research outlook in section\ref{sec:4}.

\section{Model updates}\label{sec:2}

It is well known that the UrQMD model was initially designed to simulate HICs in the energy range from SIS to
RHIC, in which the contribution of nuclear mean-field potentials to the dynamics of the reaction
is considered to be weak. However, for a more systematic and delicate investigation of the HICs in a broader energy region, this model has been successfully extended to describe
HICs with beam energies from as low as several tens of MeV/nucleon (low SIS) up to the highest energy available at
CERN LHC \cite{Bass:1998ca,Bleicher:1999xi,Petersen:2006vm,Li:2011zzp,Li:2012ta,Li:2016wkb}. At low beam energies, this model is based on the same principle as the quantum molecular dynamics model, in which a particle is represented by a Gaussian wave packet of a certain width. The equations of motion for the coordination and momentum of the {\it i}th nucleon, $\textbf{r}_{i}$ and $\textbf{p}_{i}$, read as follows:
\begin{eqnarray}
\dot{\textbf{r}}_{i}=\frac{\partial H}{\partial\textbf{ p}_{i}},
\dot{\textbf{p}}_{i}=-\frac{\partial H}{\partial \textbf{r}_{i}}.
\end{eqnarray}
Here, {\it H} is the Hamiltonian function of the system, which comprises the kinetic energy and the effective interaction potential energy. In recent years, to better describe the recent experimental data for HICs at SIS energies, the Skyrme potential energy density functional has been introduced into the nuclear potential of the UrQMD (the SV-mas08 interaction that yields the incompressibility $K_0$ = 233 MeV is employed in this current work), and it has been found that when properly set at the in-medium nucleon-nucleon cross section, some recent published experimental data, especially the collective flows of light clusters, can be reproduced fairly well. See Refs.~\cite{Wang:2013wca,wyj-sym} for more details.

The Hamiltonian of kaon mesons can be written as follows,

\begin{equation}
H_{K}=\sum_{i=1}^{N_{K}}[\omega(\textbf{p}_i,\rho_i)+V^{coul}_i].
\end{equation}
Here, the Coulomb interaction between a charged kaon and proton (kaon) is calculated in the same way as that between protons, and $\omega$ and $\textbf{p}$ are the energies and momenta of the (anti-)kaon mesons, respectively, which can be written as follows:
For $K$,
\begin{equation}
\begin{aligned}
\omega_K(\textbf{p},\rho) = \sqrt{\textbf{p}^{*2} +m_{K}^{2}- a_K\rho_s+(b_K \rho)^2 }+b_K \rho,\label{eq1}
\end{aligned}
\end{equation}
and for $\bar{K}$,
\begin{equation}
\begin{aligned}
\omega_{\bar{K}}(\textbf{p},\rho) = \sqrt{\textbf{p}^{*2} +m_{\bar{K}}^{2}- a_{\bar{K}}\rho_s+(b_K \rho)^2 }-b_K \rho. \label{eq2}
\end{aligned}
\end{equation}
Here, $\rho_s$ is the baryon scalar density and $\boldsymbol{p}^{*}=\boldsymbol{p}\mp\boldsymbol{V}$ is the
kaon effective momentum, where $V_{\mu}=\frac{3}{8f_{\pi }^{2}} j_\mu$ is the kaon
vector potential with $j_\mu$ being the baryon
four-vector current, and $f_{\pi}$ is the pseudoscalar meson decay constant. If $\boldsymbol{V}=0$, then the spatial component of the vector field vanishes \cite{Fuchs:2005zg}. The $b_{K}=3/(8f_{\pi}^{2})\approx$0.333 GeVfm$^{3}$, assuming $f_{\pi}$=93 MeV and $a_K$ and $a_{\bar{K}}$ are 0.18 and 0.31 GeV$^2$fm$^3$, respectively. With these parameters taken from Refs.~\cite{Feng:2013zya,Feng:2015yka}, the kaon and antikaon optical potentials $U_K$ and $U_{\bar{K}}$ at nuclear saturation density are about 25 MeV and -100 MeV, respectively.

The motion of a kaon meson in a nuclear medium can then be obtained as follows,
\begin{eqnarray}
   {\frac{d\textbf{r}}{dt}}  & = &  {\partial \omega_{K(\bar{K})}
                         \over \partial \textbf{p}}\nonumber\\
                         &=&\frac{{{\textbf{p}^*}}}{{\sqrt { \textbf{p}^{*2}+m_{K(\bar{K})}^2 - {a_{K(\bar{K})}}{\rho _s} + {{({b_K}\rho )}^2}} }} , \\\label{eq3}
    {\frac{d\textbf{p}}{dt}} & = & -{\partial \omega_{K(\bar{K})}
                         \over \partial \textbf{r}}-\frac{\partial V^{\textrm{Coul}}}{\partial \textbf{r}}, \label{eq4}
\end{eqnarray}
in which the kaon velocity-dependent (Lorentz-like) force originating from the spatial components $\textbf{V}$ of the
kaon vector field is included \cite{Zheng:2002mj,Feng:2013zya,Larionov:2005eb}. It is obvious that the LF force will disappear when replacing the kaon effective momentum $\boldsymbol{p}^{*}$ in Eqs.~(\ref{eq1}) and (\ref{eq2}) with $\boldsymbol{p}$.

As for the collision term, compared to other microscopic transport models that are frequently used to simulate HICs at low and intermediate energies, there are two main treatments: the unique collision time table of all individual collisions by ensuring the reference-frame independence and the {\it two-step} new particle production process. The former has been recently revisited in the Transport Simulation Code Evaluation Project \cite{Xu:2016lue,Zhang:2017esm}. With respect to the latter, new particles ($\Delta$s, pions, kaons, etc.) other than nucleons can either be produced in s-channel collisions, the decay of hadronic resonances, or the fragmentation of color strings. For more details, please see Refs.~\cite{Bass:1998ca,Bleicher:1999xi}. It has been found that the strangeness production in $pp$, $pA$ and $AA$ collisions from SIS up to RHIC energies can be described reasonably well \cite{Bratkovskaya:2004kv}. After updating new decay channels, the recently measured yields of $\phi$ and $\Xi^{-}$ data at SIS energies can be much better reproduced \cite{Steinheimer:2015sha,Steinheimer:2016vzu}.


\section{Results and Discussions}\label{sec:3}

In this work, we investigate the effect of the KN potential, particularly its Lorentz-like component, on the flow of positively charged kaon mesons produced from Au + Au collision at 1.5 GeV$/$nucleon. For each case, we calculated three million events in the considered range of impact parameters 5.9 $< $b $< $10.2 fm.

\begin{figure}[H]
\begin{centering}
\includegraphics[width=0.5\textwidth]{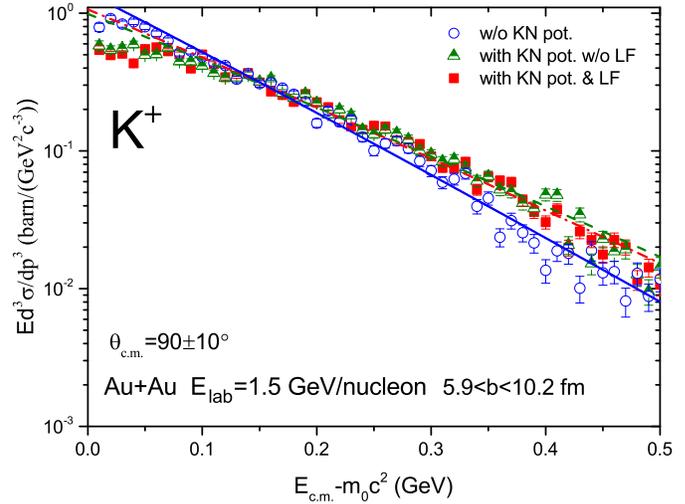}
\caption {\label{fig1} Invariant cross section as a function of the kinetic energy $E_{c.m.}-m_0c^2$ for K$^{+}$ produced from $^{197}$Au+$^{197}$Au collisions at 1.5 GeV$/$nucleon within the impact parameter range $5.9\sim10.2$ fm and $\theta_{c.m.}$=$80^{\circ}$$\sim$$100^{\circ}$. The lines represent the Maxwell-Boltzmann distribution $Ed^3\sigma/dp^3 \sim Ee^{-E/T}$ fitted to the simulations within $0.10 < E_{c.m.}-m_0c^2 <0.45$ GeV, which is analogous to the experimental window in Ref.~\cite{Forster:2007qk}.}
\end{centering}
\end{figure}

First, Figure \ref{fig1} shows the invariant cross section for K$^{+}$ as a function of its kinetic energy $E_{c.m.}-m_0c^2$. We can see that the energy distribution calculated without the KN potential is steeper than the other calculations, which leads to a smaller ``temperature'' $T$. With the Maxwell-Boltzmann distribution $Ed^3\sigma/dp^3 \sim Ee^{-E/T}$ fitted to the calculations (scattered symbols), we obtain the inverse slope parameters $T$ of $85\pm2$, $105\pm3$, and $101\pm2$ MeV for the calculations without the KN potential, with the KN potential but without the LF, and with both the KN potential and LF, respectively. We find that the extracted $T$ parameters from the experimental measurements with the same collision system and beam energy but for near-central collisions are $111\pm2$ MeV in Ref.~\cite{Forster:2007qk} and $116\pm7$ MeV in Ref.~\cite{Forster:2003vc}. Therefore, the temperature obtained from calculations without the KN potential is much smaller than it obtained from the experimental data, whereas those obtained from calculations with the KN potentials are closer to that of the experimental data. Furthermore, the influence of the LF on the energy distribution for K$^{+}$ is quite weak. Similar conclusions can be found in Refs.~\cite{Feng:2013zya,Larionov:2005eb} as well by using the LQMD model and the Boltzmann Uehling-Uhlenbeck (BUU) model, respectively.

\begin{figure}[H]
\centering
\includegraphics[width=0.40\textwidth]{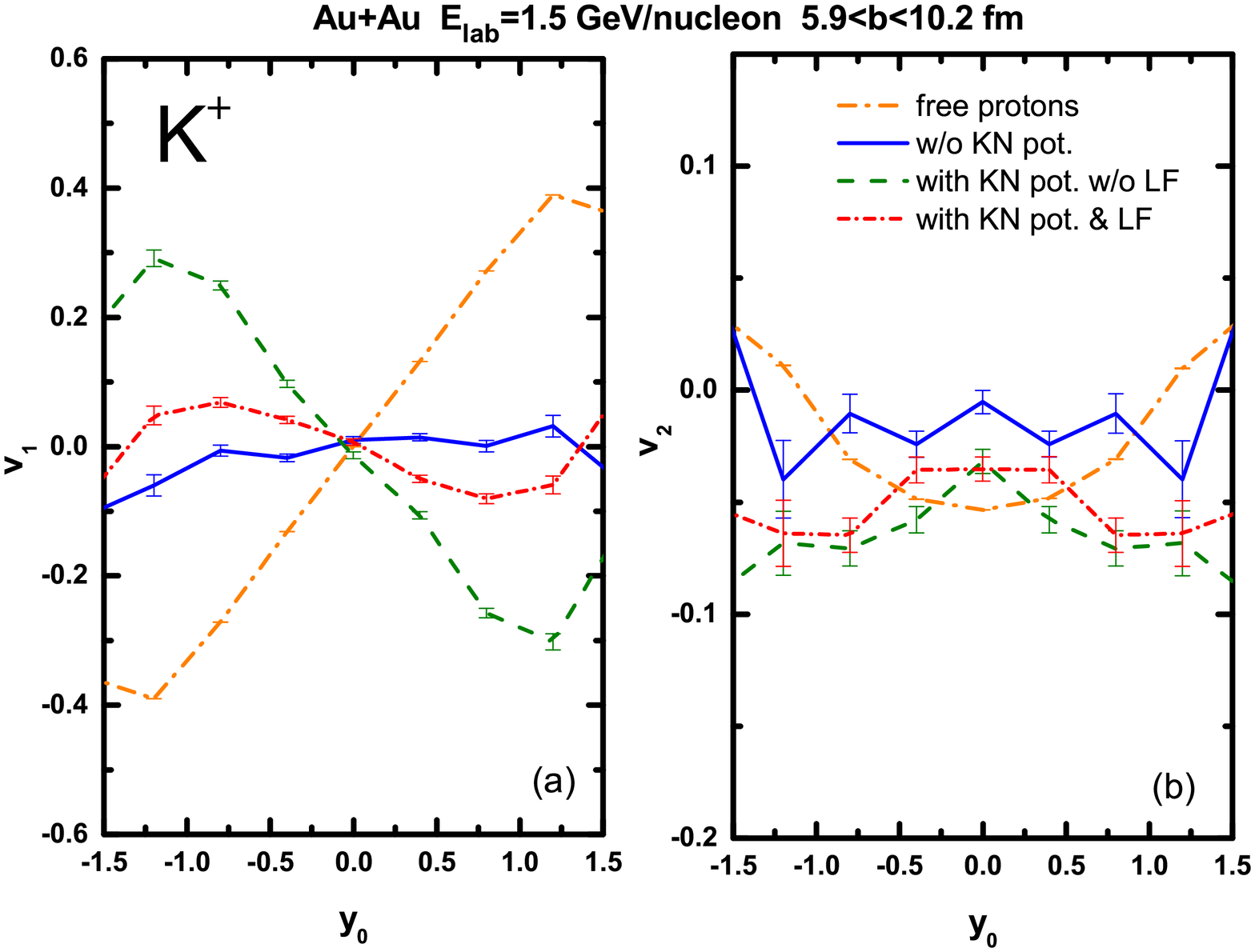}
\caption {\label{fig2} Rapidity distribution of the directed flow
$v_{1}$ (a) and the elliptic flow $v_{2}$ (b) of the K$^{+}$ mesons produced in $^{197}$Au+$^{197}$Au collisions at an incident energy of 1.5 GeV/nucleon within the impact parameter range $5.9< b <10.2$ fm. Calculations with the KN potential and without (dashed lines) or with (short-dash-dotted lines) the LF are compared to those without the KN potential (solid lines). The flows of free protons are also shown by the dash-dotted lines. }
\end{figure}

Figure \ref{fig2} shows the directed flow ($v_1=\langle\frac{p_x}{p_t}\rangle$, where $p_t=\sqrt{p_x^2+p_y^2}$) and the elliptic flow
($v_2=\langle\frac{p_x^2-p_y^2}{p_t}\rangle$) of $K^+$ as a function of the normalized rapidity $y_{0}=y_{z}/y_{pro}$. Here, $p_x$ and $p_y$ are two transverse components of the momentum, $y_z=\frac{1}{2}\ln\frac{E+p_z}{E-p_z}$ is the longitudinal rapidity, and $y_{pro}$ is the projectile rapidity in the center-of-mass system.

In Figure \ref{fig2}(a), we can see that the slope of $v_{1}$ at mid-rapidity without consideration of the KN potential (solid line) is slightly positive (since only the Coulomb potential is considered for kaons in this case), whereas it becomes visibly negative when the KN potential is taken into account but without consideration of the LF (dashed line). This is because the static KN potential is strongly repulsive so that the direction of the kaon motion tends to be opposite to that of nucleon motion. At this beam energy, the slope of the directed flow of nucleons at mid-rapidity is positive, as seen by the dash-dotted line, so we can see the anti-flow of K$^{+}$. This repulsion effect can also be seen in the elliptic flow shown in Figure \ref{fig2}(b), where the $v_{2}$ of K$^{+}$ calculated with the static KN potential is much more negative than that without the KN potential, which indicates a pronounced out-of-plane K$^{+}$ squeezeout. When the LF is also taken into account, both the $v_{1}$ slope value at mid-rapidity and the absolute value of $v_{2}$ of K$^{+}$ mesons (each shown by the short-dash-dotted line) become smaller than those considering only the static KN potential. This is because the LF for positively charged kaons provides the attractive potential to pull the kaons back to the nucleons. We note that a negative flow for K$^{+}$ was observed by the FOPI collaboration in the collision with a smaller system $^{58}$Ni+$^{58}$Ni at a similar beam energy 1.91 GeV$/$nucleon \cite{Zinyuk:2014zor}, as well as by some model calculations \cite{Zheng:2002mj,Feng:2013zya,Larionov:2005eb}.


\begin{figure}[H]
\begin{centering}
\includegraphics[width=0.4\textwidth]{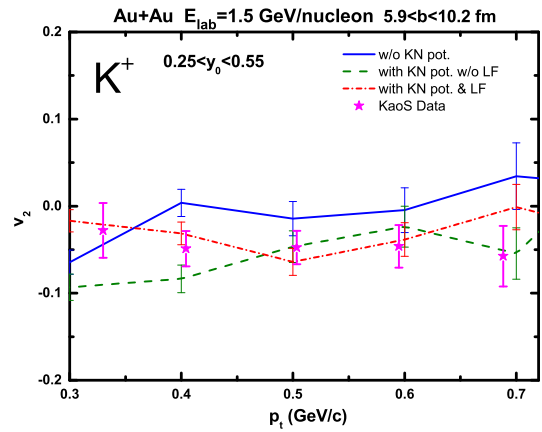}
\caption {\label{fig3}Transverse momentum $p_t$ distribution of the
elliptic flow $v_2$ of K$^{+}$ mesons produced in $^{197}$Au+$^{197}$Au collisions at 1.5 GeV$/$nucleon within the impact parameter range $5.9\sim10.2$ fm and the rapidity cut $0.25\sim0.55$. The KaoS experimental data are taken from Ref.~\cite{Ha06} (stars) and compared to calculations without and with KN potentials (distinguished by different lines). }
\end{centering}
\end{figure}

We can further examine the transverse momentum $p_t$ dependence of the elliptic flow $v_2$ of K$^{+}$ in Figure \ref{fig3}, in which the KaoS data \cite{Ha06} are shown for comparison. We can see a weak dependence on both the $p_t$ and the KN potential although with the existence of large statistical errors. More specifically, the values of $v_2$ calculated without the KN potential are somewhat higher than those of the two cases with the KN potential, as well as the experimental data. The weak sensitivity of the elliptic flow of K$^{+}$ as functions of both rapidity and transverse momentum in the KN potential was also found by IQMD model calculations in Ref.~\cite{Ha06}.

\begin{figure}[H]
\begin{centering}
\includegraphics[width=0.5\textwidth]{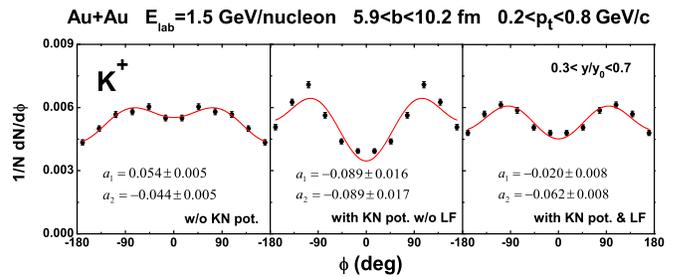}
\caption {\label{fig4}Azimuthal distribution of K$^{+}$ mesons in $^{197}$Au+$^{197}$Au
collisions at 1.5 GeV/nucleon within the impact parameter range from $5.9\sim10.2$ fm. The $y_0$ cut $0.3\sim0.7$ and the $p_t$ cut $0.2\sim0.8$ GeV/c are employed. The lines represent fits to the calculated results (scattered solid squares) assuming $\frac{dN}{d\phi}\cong1+2a_{1}cos(\phi)+2a_{2}cos(2\phi)$, and the corresponding $a_1$ and $a_2$ values are also given in each plot.}
\end{centering}
\end{figure}

\begin{figure}[H]
\begin{centering}
\includegraphics[width=0.5\textwidth]{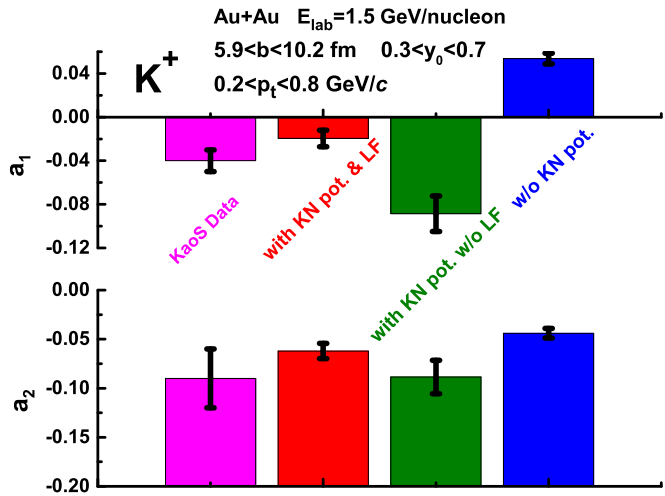}
\caption {\label{fig5}Comparison of the calculated $a_{1}$ (upper plot) and $a_2$ (lower plot) values with the KaoS experimental data (leftmost bars) taken from Ref.~\cite{kaos}.}
\end{centering}
\end{figure}

Figure \ref{fig4} shows the azimuthal distributions of K$^{+}$ in the rapidity bin $0.3< y_0<0.7$ without (left) and with (right) the KN potential as well as with only a static KN potential (middle). We used a $p_t$ cut of $0.2\sim0.8$ GeV/c for all cases. To evaluate the anisotropy of the distributions, we used the form $\frac{dN}{d\phi}\cong1+2a_{1}cos(\phi)+2a_{2}cos(2\phi)$ in the same way as in the KaoS experiment \cite{kaos}. We indicate the fitting results with lines and also give the extracted $a_1$ and $a_2$ values for each plot. It is obvious that the $a_1$ and $a_2$ parameters have the same meaning as the $v_1$ and $v_2$ flow parameters although they are not quantitatively equivalent; the negative (positive) $a_1$ value represents a negative (positive)-directed flow, and the negative (positive) $a_2$ value represents an out-of-plane (in-plane) emission. We can see that the values of both $a_1$ and $a_2$ with the full KN potential lie between the other two cases. Furthermore, in Figure \ref{fig5}, we show a comparison of the $a_{1}$ and $a_{2}$ values with the corresponding KaoS experimental data \cite{kaos} with the same conditions as those in Figure \ref{fig4}. We note that the $a_1$ data presented in Ref.~\cite{kaos} is positive but we use its reverse value here since the relative position of the projectile with respect to the target in our calculations is contrary to that in the experiment. Again, we can obviously see that the directed flow is more sensitively influenced by the KN potential than the elliptic flow since the $a_1$ value changes more drastically (and even changes sign) than $a_2$ when the KN potentials are taken into account. The cancellation effect between the scalar and vector aspects of the KN potential is also essential since the calculated $a_1$ value with the KN potential but without the LF is more negative and moves away from the data. Finally, both the $a_1$ and $a_2$ calculated results match the KaoS data well within errors if the KN potential including the LF is used, whereas neither can reproduce the data if the KN potential is not considered.

\section{Summary and Outlook}\label{sec:4}
In summary, within the updated version of the UrQMD transport model, in which we introduce the kaon¨Cnucleon (KN) potential, including both the static and the kaon velocity-dependent (Lorentz-like) force originating from the spatial components of the kaon vector field, we investigated the collective flow of K$^{+}$ produced by $^{197}$Au+$^{197}$Au collisions at the incident energy of 1.5 GeV/nucleon. We found that the directed flow of K$^{+}$ mesons is more visibly affected by the KN potential than the elliptic flow. Due to the repulsive nature of the static KN potential, we observed negative-directed and elliptic flow parameters. Furthermore, the attractive Lorentz-like force largely cancels the negative flows (especially the directed flow) so that both the directed and elliptic flow data of the KaoS experiment can be well reproduced.

Due to the current large uncertainty in the study of $\bar{K}$N potential, a further numerical calculation relevant to this topic is in progress. Moreover, due to the large absorption effect of the $\bar{K}$ mesons in the nuclear medium, the related two-body scattering process will be dealt with more consistently.

\Acknowledgements{The authors acknowledge the support of the computing server C3S2 at the Huzhou University. This work is supported in part by the National Natural
Science Foundation of China (Nos. 11375062, 11505057, and 11647306), and the Zhejiang Provincial Natural Science Foundation of China (No. LY18A050002), and the project is sponsored
by SRF for ROCS, SEM.}

\InterestConflict{The authors declare that they have no conflict of interest.}



\end{multicols}
\end{document}